\documentclass[superscriptaddress,twocolumn,showpacs,preprintnumbers,amsmath,amssymb,prb,aps]{revtex4}
\usepackage[all]{xy}

\usepackage{graphicx,psfrag,times,epsfig,color}
\usepackage{verbatim,natbib}

\usepackage{makeidx}
\usepackage{amsmath}
\usepackage{bm}
\usepackage{amsfonts}
\usepackage{amssymb}
\usepackage{hyperref}

\begin{document}

%\title{Superconductivity of non-symmetric $s-p$ hybrid diatomic linear chain}
\title{Applying experimental constraints to a one-dimensional model for ${\rm BiS_2}$ superconductivity}
\author{M.~A.~Griffith}
\email[Corresponding author: ]{griffith_mas@hotmail.com}
\affiliation{Centro Brasileiro de Pesquisas F\'isicas, Rua Dr. Xavier Sigaud 150, Urca, 22290-180 Rio de Janeiro, RJ, Brazil}
\author{K.~Foyevtsova}
\affiliation{Department of Physics \& Astronomy, University of British Columbia, Vancouver, British Columbia V6T 1Z1, Canada}
\affiliation{Quantum Matter Institute, University of British Columbia, Vancouver, British Columbia V6T 1Z4, Canada}
\author{M.~A.~Continentino}
\affiliation{Centro Brasileiro de Pesquisas F\'isicas, Rua Dr. Xavier Sigaud 150, Urca, 22290-180 Rio de Janeiro, RJ, Brazil}
\author{G.~B.~Martins}
\email[Corresponding author: ]{martins@oakland.edu}
\affiliation{Department of Physics, Oakland University, Rochester, MI 48309, USA}

\begin{abstract}
{Recent ARPES measurements [Phys. Rev. B {\bf 92}, 041113 (2015)] have confirmed the one-dimensional character of 
the electronic structure of ${\rm CeO_{0.5}F_{0.5}BiS_2}$, a representative 
of ${\rm BiS_2}$-based superconductors. In addition, several 
members of this family present sizable increase in the superconducting transition temperature $T_c$ under 
application of hydrostatic pressure. Motivated by these two results, we propose 
a one-dimensional three-orbital model, whose kinetic energy part,  obtained through {\it ab initio} 
calculations, is supplemented by pair-scattering terms, which are 
treated at the mean-field level. 
We solve the gap equations self-consistently and then systematically probe which combination of pair-scattering terms 
gives results consistent with experiment, namely, a superconducting dome with a 
maximum $T_c$ at the right chemical potential and a sizable increase in $T_c$ when 
the magnitude of the hoppings is increased. For these constraints to be satisfied multi-gap superconductivity  
is required, in agreement with experiments, and one of the hoppings has a dominant influence over the 
increase of $T_c$ with pressure. 
}
\end{abstract}
\pacs{74.20.Mn,74.20.Rp,74.70.-b}
\maketitle

\newcommand{\h}{\mathcal{H}}
\setcounter{MaxMatrixCols}{30}

{\it Introduction.}
After the discovery of the cuprates in 1986 \cite{Bednorz1986}, the search for new layered superconducting materials
has attracted much attention, with important discoveries occurring in the last 15 years. For example, it was discovered in
2001 that ${\rm MgB_2}$ has $T_c=39$ K \cite{Nagamatsu2001} and in 2008 superconductivity (SC) in the iron
pnictides was reported \cite{Kamihara2008}.
Both ${\rm MgB_2}$ and the iron pnictides have highlighted the importance of multiband SC 
\cite{Lin2014}, to the point that
the recent literature on cuprates devoted to multiband models has substantially increased \cite{White2015}. An unrelated
development has been the explosion of research in topological superconductors \cite{Beenakker2013}, due to proposals
to `engineer' Majorana fermion quasiparticles through midgap excitations of a chiral
p-wave superconductor. This has led to renewed interest in the Ruthenate compound ${\rm Sr_2RuO_4}$, discovered in 1994
\cite{Maeno1994}, which is one of the few candidates to realizing p-wave-type SC
\cite{Mackenzie2003,Kallin2012}, another candidate being the organic superconductor ${\rm (TMTSF)_2PF_6}$.
It should also be emphasized that, as was the case for intermetallics with A15 structure (like ${\rm Nb_3Sn}$
or ${\rm V_3Si}$) \cite{Bok2012}, the Ruthenates display `hidden' quasi-one-dimensional (quasi-1d) 
SC \cite{Raghu2010} (while organic superconductors are explicitly 1d). Finally, we also mention 
SC in doped semiconductors, studied since before the 60s \cite{Cohen1964}, 
with the interest greatly increasing after the discovery of SC 
in Boron-doped Diamond with $T_c=4$ K \cite{Sidorov2010}. 

It is then interesting that one of the latest families of layered superconductors to be 
discovered, those containing ${\rm BiS_2}$ planes, \cite{Mizuguchi2012b} presents many of the characteristics mentioned 
above: a layered structure, similar to cuprates and pnictides \cite{Mizuguchi2014}; 
a double superconducting gap as in ${\rm MgB_2}$ \cite{Liu2014a}; its minimal model 
contains two bands \cite{Usui2012a}, and Fermi surface nesting effects seem 
to be important \cite{Martins2013} (as in the iron pnictides); because it contains a heavy element (Bismuth), spin-orbit effects are 
enhanced and some proposals linking ${\rm BiS_2}$ to spin-triplet pairing and a weak topological superconducting state 
have been made \cite{Yang2013}; based on first-principles electronic structure calculations, it has been pointed 
out the `subtle' 1d character of its band structure \cite{Usui2012a}, which has been recently 
confirmed experimentally through polarization-dependent Angular Resolved Photoemission Spectroscopy 
(ARPES) measurements \cite{Sugimoto2015a}; finally, a few members 
of the ${\rm BiS_2}$ family have semiconducting parent compounds that become metallic/superconducting with electron doping or 
application of moderate hydrostatic pressure, which also can lead to sizable increase in ${\rm T_c}$ \cite{Wolowiec2013a}.

In this work, to advance the understanding of SC 
in ${\rm BiS_2}$, where there is no consensus yet if it is of the conventional or unconventional type \cite{Hirsch2015}, 
we concentrate in these last two aspects: one-dimensionality of the electronic structure 
and the pronounced effects pressure has over the superconducting phase. To model that, the authors take the following approach: 
i) adopt a 1d three-orbital model for ${\rm BiS_2}$, adding the Cooper-pairing by hand, ii) solve 
the gap equations at the mean-field level, iii) study the dependence of the superconducting gap 
with the variation of the hopping terms, whose magnitude one expects to increase under applied pressure iv) 
decide on the acceptance or not of specific pair-scattering terms based on semi-quantitative agreement with 
experiments. Regarding this last point, we look specifically in what range of electron-filling a 
superconducting dome is obtained (see Fig.~3) and how SC varies with hopping parameters. 
To make the connection with ${\rm BiS_2}$ more explicit, and thus obtain semi-quantitative 
agreement with experiments, all the parameter values of the single-particle Hamiltonian 
were obtained through first-principles Density Functional Theory (DFT) calculations for a 
two-dimensional (2d) five-band model (see Table \ref{tab:hopp}).

We can summarize our results as follows: Taking into account a three-orbital model, 
where Sulfur contributes with orbitals $s$ and $p$, and Bismuth with a $p$ orbital (see Fig.~1), 
we considered all possible pair-scattering terms (intra and interband, restricted to pairs formed 
by same-band electrons), individually and in conjunction, and solved the resulting gap equations at the 
mean-field level. We obtain that i) no single-band pair-scattering process, acting isolatedly, can describe the experiments 
(as specifically defined above), unless an unrealistic coupling is assumed ($g > 0.1$ eV); this seems to indicate that 
multi-gap SC is a natural consequence of our model ii) two different types of 
multi-gap SC (see detailed description below) are in semi-quantitative agreement with experiments 
iii) the gap dependence with hopping (see Fig.~4) indicates a qualitative difference 
between the two hoppings considered in our model.  
These important results establish an appropriate effective 1d model to simulate the properties of 
${\rm BiS_2}$. We expect that our work will motivate other groups to investigate other similar purely 
1d effective models. 

\begin{figure}
\centering
\includegraphics[width =3.4in]{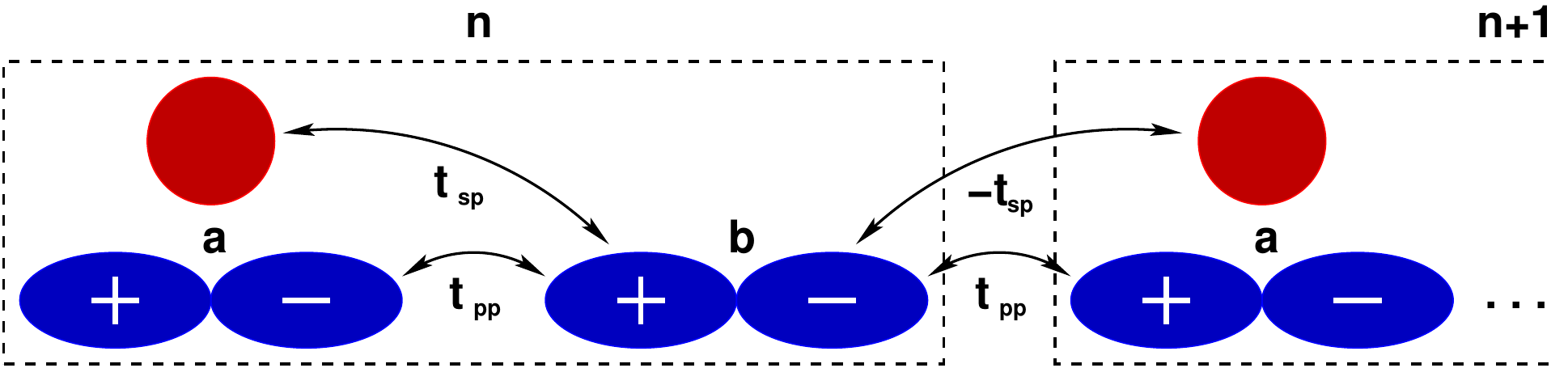}
\caption{1d model for ${\rm BiS_2}$. $t_{pp}$ and $t_{sp}$ are the hopping terms
	considered in our model. The dashed box indicates the unit cell with atoms $a$ (Sulfur, 
	with orbitals $s$ and $p$) and $b$ (Bismuth, with just one $p$ orbital). 
Notice the alternating signs of the $t_{sp}$ hoppings  \cite{Continentino2014}.
}
\label{figure1}
\end{figure}

{\it Model.}
We consider a linear chain with a unit cell consisting of two sites denoted $a$ and $b$,
see Fig.~\ref{figure1}.
The $a$ sites (Sulfur) have orbitals $s$ and $p$, while $b$ sites (Bismuth) have just one $p$ orbital.
In second quantization notation, the annihilation operator
for an $s$ orbital in unit cell $n$ is denoted as $c_n$, and those for Sulfur and Bismuth $p$ orbitals are denoted
$p_{a,n}$ and $p_{b,n}$, respectively. 
The non-interacting part of the Hamiltonian can then be written as
\begin{eqnarray}
	\mathcal{H}&=&\sum_{n}\{(\epsilon_{s,n}+\mu)n_{s,n}+\sum_{i=a,b}(\epsilon_{p_{i},n}+\mu)n_{p_{i},n}  \nonumber \\
&+&  t_{pp} [p^{\dag}_{a,n}p_{b,n} + p^{\dag}_{b,n}p_{a,n+1}+{\rm h.c.}]  \nonumber \\
&+&  t_{sp}[c^{\dag}_{n}p_{b,n} - p^{\dag}_{b,n}c_{n+1}+{\rm h.c.}] \}
\label{hamiltsp}
\end{eqnarray}
where $\epsilon_{s,n}$ and $\epsilon_{p_{i},n}$ describe the energy levels of orbitals $s$ and $p$ (for site
$i=a,b$) at unit cell $n$, respectively; $n_{s,n}=c^{\dag}_{n}c_{n}$ and $n_{p_{i},n}=p^{\dag}_{i,n}p_{i,n}$
are the number operators, and $\mu$ is the chemical potential. 
The hopping parameters are indicated in Fig.~\ref{figure1} and the values used in this work 
(along with orbital energies and chemical potential) are listed in Table \ref{tab:hopp} in eV units. Note that 
these parameter values were obtained through a full DFT calculation. 
The hoppings kept for the 1d model here studied were all the nearest neighbor hoppings in excess of $0.5$ eV. 

\begin{table}
\caption{Partial list of tight-binding parameters (in eV) for the 2d five-orbital model. 
The same parameters are used for the 1d three-orbital model depicted in Fig.~\ref{figure1}. 
The chemical potential corresponds to 1/8-filling of the $p_b$ orbital in the 1d model.\label{tab:hopp}}
\vspace{0.3cm}
 \begin{tabular}{|cccccc|}\hline
	 $\epsilon_{s,n}$ & $\epsilon_{p_{a},n}$ & $\epsilon_{p_{b},n}$ & $t_{sp}$ & $t_{pp}$ & $\mu$ \\
\hline
$-11.2840$   & $-1.2691$ & $0.1635$  & $-0.9952$ & $-0.8155$ & $0.5007$ \\ 
\hline
 \end{tabular}
\end{table}

\begin{figure}
\centering
\includegraphics[trim=0.0cm 10.5cm 12.5cm 0.0cm, width =4.0in]{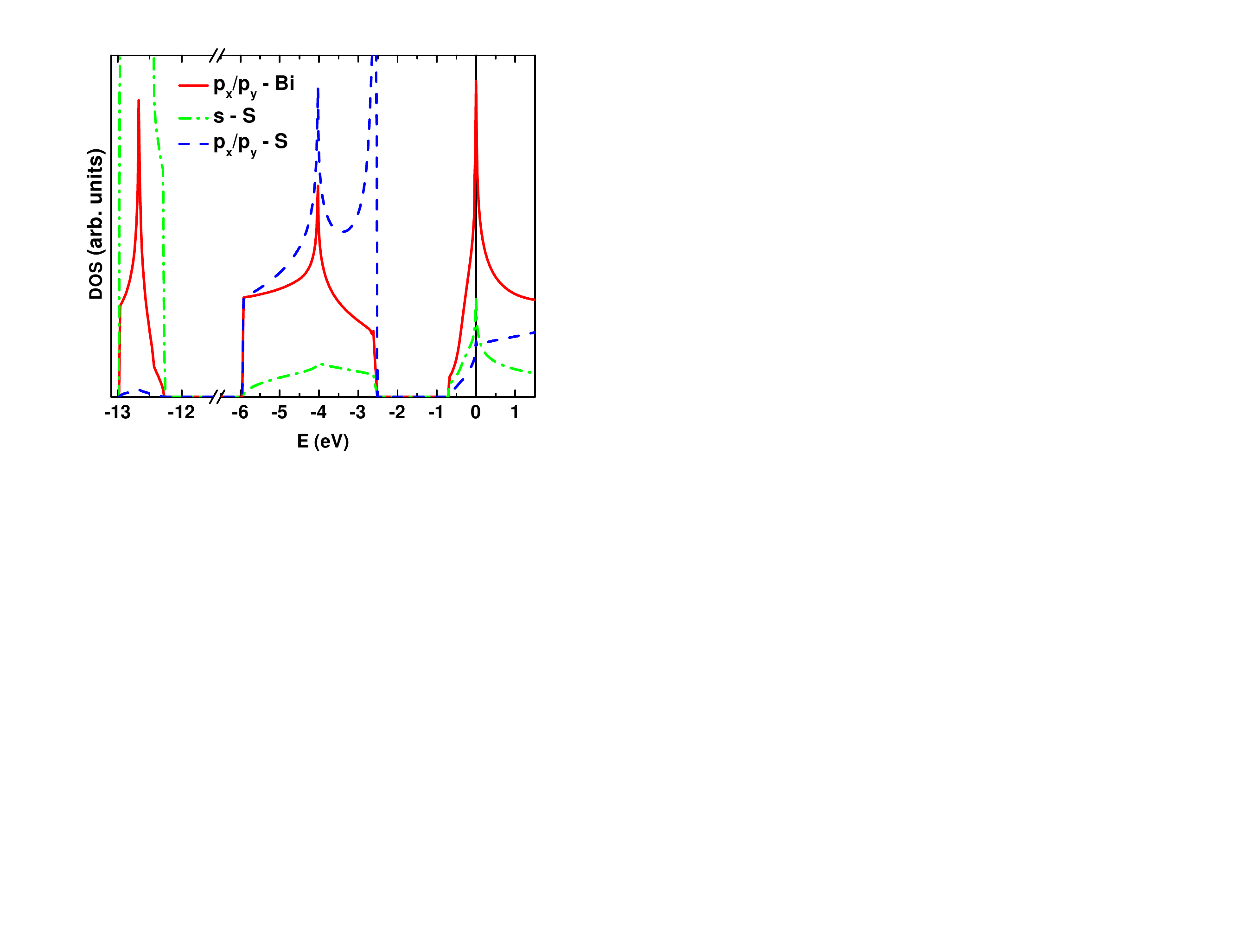}
\caption{Density of states obtained through DFT for the five-orbital model of 
the ${\rm LaOBiS_{2}}$ compound. Note the importance of the Sulfur
$s$-orbitals [dot-dashed (green) line] at the Fermi energy ($E_F=0$), justifying its inclusion in the model
described in Fig.~\ref{figure1}. In addition, the sequence of van Hove singularities 
at the bottom of the conduction band and at the top of the valence band 
indicates the quasi-1d character of the electronic structure. 
}
\label{figure2}
\end{figure}

An early {\it 2d} minimal model for ${\rm BiS_2}$ contains two orbitals: Bismuth $p_x$ 
and $p_y$ orbitals \cite{Usui2012a}. Therefore, before deriving the self-consistent gap equations, 
the inclusion of the Sulfur $p$ and $s$ orbitals 
should be justified, mainly the latter one, which lies deep below the Fermi energy (see DFT parameter 
values in Table \ref{tab:hopp}). Figure~\ref{figure2} shows the density of states (DOS) obtained for 
a {\it 2d} model of ${\rm BiS_2}$ involving five orbitals: two Bismuth $p$ orbitals ($p_x$ and $p_y$), 
two Sulfur $p$ orbitals ($p_x$ and $p_y$), and one Sulfur $s$ orbital. From the 
examination of the DOS one can conclude that, at the 
Fermi energy $E_F=0$ (which, in this plot, is between 1/8- and 1/4-filling, for the 2d model), the 
participation of the Sulfur $s$ orbital [dot-dashed (green) curve], is quite relevant, 
even more than that of the Sulfur $p$ orbitals [dashed (blue) curve]. In addition, it is easy to recognize the characteristic 
1d DOS profile for the Sulfur $p$ orbital at the top of the valence band and for the Bismuth 
$p$ and Sulfur $s$ orbitals at the bottom of the conduction band. This, coupled to the above mentioned 
polarized ARPES results indicating the one-dimensionality of the electronic structure of ${\rm BiS_2}$, 
justifies our model. We now proceed to the derivation of the self-consistent gap equations. 

{\it Self-consistent gap equations at zero temperature.}
After taking a Fourier transform of the non-interacting part, and introducing pair-scattering terms between the electrons, 
the total Hamiltonian can be written as
\begin{eqnarray}
\mathcal{H}(k)&=&\sum_{k}\{(\epsilon_{as}+\mu)c^{\dag}_{k}c_{k}+\sum_{i=a,b}(\epsilon_{p_{i}}+\mu)p^{\dag}_{ik}p_{ik} \nonumber \\
&+& 2t_{pp}\cos(k)[p^{\dag}_{ak}p_{bk}+{\rm h.c.} ] \nonumber \\
&+& 2it_{sp}\sin(k)[c^{\dag}_{k}p_{bk} - {\rm h.c.}] \nonumber \\
&-& \sum_{i,j,k,k^{\prime}} g_{ij}[\gamma^{\dagger}_{i,k+}\gamma^{\dagger}_{i,\bar{k}-}\gamma_{j,k^{\prime}+}\gamma_{j,\bar{k^{\prime}}-}+{\rm h.c.}]\}
\end{eqnarray}
where, in the last line, $\gamma_{i/j,k\sigma}$ ($\sigma=\pm$ and $\bar{k}$ indicates $-k$) 
stands for either one of $c_{k\sigma}$, $p_{a,k\sigma}$, or $p_{b,k\sigma}$. Note that 
it is implicit in the form of the expression for the pair-scattering term that we are only 
considering Cooper pairs composed of electrons from the same band, 
as pairing of different-band electrons tends to promote pair-density-waive (inhomogeneous) 
superconducting ground states \cite{Caldas2012}. 
Already anticipating results that will be discussed below (see Fig.~3), we describe 
how to obtain the gap equations when an specific set of pair-scattering processes are taken in account. Considering 
terms involving {\it intraband} scattering in the $s$ and $p_b$ bands and {\it interband} scattering between the 
$s$ and $p_b$ bands, the last line of eq.~(2) (which we denote as $\Delta^{SC}$) can be written as
\begin{eqnarray}
\Delta^{SC}&=&-\sum_{k}\{ g_{ss}[ c_{k+}^{\dag}c_{\bar{k}-}^{\dag}c_{\bar{k}-}c_{k+}]\nonumber \\
&+&g_{p_{b}p_{b}}[ p_{b,k+}^{\dag}p_{b,\bar{k}-}^{\dag}p_{p,\bar{k}-}p_{b,k+}]\nonumber \\
&+&g_{sp_{b}} [ c_{k+}^{\dag}c_{\bar{k}-}^{\dag}p_{b,\bar{k}-}p_{b,k+}\nonumber \\
&+& p_{b,k+}^{\dag}p_{b,\bar{k}-}^{\dag}c_{\bar{k}-}c_{k+} ]\}.
\end{eqnarray}

Note that, for simplicity, we consider the pairing couplings $g_{ij}$ 
as being $k$-independent, {\it i.e.}, we assume s-wave pairing functions. 
Applying a mean-field decoupling eq.~(3) becomes 
\begin{eqnarray}
\Delta^{SC}&=&-\sum_{k} \{\Delta_{ss}(c_{\bar{k}-}c_{k+} + c_{k+}^{\dag}c_{\bar{k}-}^{\dag})\nonumber \\
&+& \Delta_{p_{b}p_{b}}(p_{p,\bar{k}-}p_{b,k+}+p_{b,k+}^{\dag}p_{b,\bar{k}-}^{\dag})\nonumber \\
&+&[\Delta_{ss}^{\prime}(p_{b,\bar{k}-}p_{b,k+} + p_{b,k+}^{\dag}p_{b,\bar{k}-}^{\dag} )\nonumber \\
&+& \Delta_{p_{b}p_{b}}^{\prime}(c_{k+}^{\dag}c_{\bar{k}-}^{\dag}+c_{\bar{k}-}c_{k+} ) ]\},
\end{eqnarray}
which can be rewritten as 
\begin{eqnarray}
\Delta^{SC}&=&-\sum_{k} \Delta_{1}(c_{\bar{k}-}c_{k+} + c_{k+}^{\dag}c_{\bar{k}-}^{\dag})\nonumber \\
&+&\Delta_{2}(p_{p,\bar{k}-}p_{b,k+}+p_{b,k+}^{\dag}p_{b,\bar{k}-}^{\dag}),
\end{eqnarray}
with the following definitions 
\begin{eqnarray}
\Delta_{1}&=& \Delta_{ss}+ \Delta^{\prime}_{p_{b}p_{b}} \nonumber \\
		 &=& g_{ss}\sum_{k}\langle c_{\bar{k}-}c_{k+} \rangle +  g_{sp_{b}}\sum_{k}\langle p_{b,\bar{k}-}p_{b,k+} \rangle\nonumber
\end{eqnarray}
and 
\begin{eqnarray}
\Delta_{2}&=& \Delta^{\prime}_{ss}+ \Delta_{p_{b}p_{b}} \nonumber \\
&=& g_{sp_{b}}\sum_{k}\langle c_{\bar{k}-}c_{k+} \rangle +  g_{p_{b}p_{b}}\sum_{k}\langle p_{b,\bar{k}-}p_{b,k+} \rangle\nonumber,
\end{eqnarray}
where $<>$ indicates an average over the ground state.
For simplicity, if we consider the following relations, $ g_{ss} = g_{p_{b}p_{b}} = g_{sp_{b}} = g $ and 
$ \Delta_{ss} = \Delta_{ss} = \Delta^{\prime}_{p_{b}p_{b}} = \Delta^{\prime}_{p_{b}p_{b}} = \Delta $, we obtain the 
gap equation as 
\begin{eqnarray}
 2\Delta = g\sum_{k}(\langle c_{\bar{k}-}c_{k+} \rangle + \langle p_{b,\bar{k}-}p_{b,k+} \rangle ).
\end{eqnarray}

We want to derive a self-consistent equation for $\Delta$ and then analyze the effect
of variations in the hopping parameters over it.
In order to determine the correlations $\langle c_{\bar{k}-}c_{k+} \rangle$ and $\langle p_{b,\bar{k}-}p_{b,k+} \rangle$ in  
the gap equation, we need to calculate the anomalous Green's functions
$\langle\langle c_{k+};c_{\bar{k}-} \rangle\rangle$ and $\langle\langle p_{b,k+};p_{b,\bar{k}-} \rangle\rangle$.
These calculations are long and tedious, and thus are presented in the supplemental material\cite{supplemental}. 
After writing the equation of motion for the propagators $\langle\langle p_{b,k+};p_{b,\bar{k}-}\rangle\rangle$ 
and $\langle\langle c_{k+};c_{\bar{k}-}\rangle\rangle$, and through lengthy algebraic manipulations, 
we arrive at expressions for $\Delta_1$ and $\Delta_2$
\begin{eqnarray}
	\Delta_{1;2}&=& -g_{ss;sp_{b}}k_{f} \pi \int^{1}_{-1} d \tilde{k} \sum^{3}_{j=1} \frac{ |D_{ss}(-\omega_{\tilde{k},j})| }{ \omega_{\tilde{k},j}r_{j} } \nonumber \\
&-&  g_{sp_{b};p_{b}p_{b}}k_{f} \pi \int^{1}_{-1} d \tilde{k} \sum^{3}_{j=1} \frac{ |D_{p_{b}p_{b}}(-\omega_{\tilde{k},j})| }{ \omega_{\tilde{k},j}r_{j} }
\end{eqnarray}
which, after the simplifying step mentioned above, results in $\Delta_1=\Delta_2=\Delta$ (the terms under the two 
integrals are fully developed in the supplemental material\cite{supplemental}).

\begin{figure}
\centering
\includegraphics[width=3.5in]{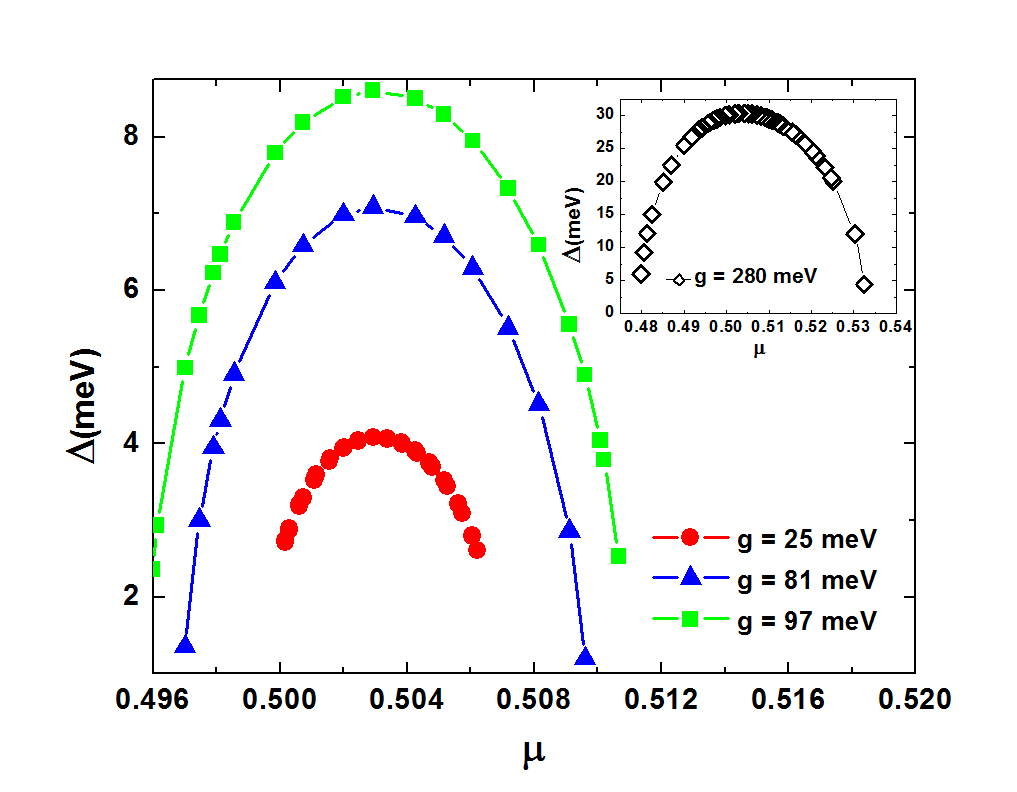}
\caption{Pairing interaction $\Delta$ as a function of the chemical potential $\mu$ for
different values of the coupling constant $g$. These results were obtained for the 
tight-binding parameter values listed in Table \ref{tab:hopp}. 
The value of $\mu$ at the center of the dome corresponds to an electron filling close to the one in $BiS_2$ compounds 
where SC has been observed (1/8-filling). The inset shows results for a larger value of $g$, 
which stabilizes the superconducting phase in a much broader interval of $\mu$. 
}
\end{figure}

% -----------------------------------------------------------------------------

{\it Results.}
As mentioned in the Introduction, our strategy was to solve the gap equations at the mean-field level 
(hopping amplitudes fixed at the values obtained by DFT), 
and look for solutions at least qualitatively compatible with experiments, 
i.e., for chemical potential values around 1/8-filling and for coupling strengths $g$ that are not 
unrealistically large. 
Taking in account the pair-scattering terms in eq.~(4) and following the derivations up to eq.~(7), 
we obtain the gap function $\Delta$, which has a dependence with $\mu$ as shown in Fig.~3, 
for three different values of coupling $g$. It is interesting to note that the value of $\mu$ 
around which the three domes are centered corresponds to an electron filling close to that where 
SC has been found for most members of the BiS2 family, i.e., 1/8-filling \cite{note1}. 
This is an important result, as $\mu$ was {\it not} fixed from the start. It is taken as a free parameter, whose 
value, obtained self-consistently, was used to determine which gap equations (for specific pair-scattering terms) 
produced acceptable results. Indeed, if the value of $\mu$ for which SC was found is too 
far removed from 1/8-filling, that gap equation (and the pair-scattering term generating it) is rejected. 

If one takes the maximum value obtained for $\Delta$ in Fig.~3 for $g=25$ meV [(red) circles], 
$\Delta \approx 4$ meV, and uses the BCS relation $2\Delta/k_BT_c=3.52$, one obtains $T_c \approx 26$ K. 
A maximum $T_c \approx 11$ K has been found for $\rm LaO_{1-x}F_xBiS_2$ at 1/8-filling ($x=0.5$) \cite{Yazici2015}, 
indicating that our results, for a realistic value of $g$, produce a $T_c$ qualitatively similar 
to experiments. 
A comment should be made on the horizontal width of the dome for the (red) solid circles curve in Fig.~3. 
At the base of the dome, the electron filling varies 
roughly from $0.25$ to $0.26$ electrons per $p_b$-orbital (Bismuth). Although there 
is still some controversy about the actual filling around which SC occurs \cite{note1}, 
a few of the published $T_c$ vs.~doping results indicate a broader dome. We believe that the narrower dome 
we obtain is an artifact of the 1d model. Indeed, the DOS close to 1/8-filling for our 1d model (not shown) 
has a very pronounced 
van Hove singularity, therefore a very strong variation of DOS with the chemical potential. 
This strong dependence, for smaller values of $g$ (as the ones plotted in the main panel in Fig.~3), 
seems to result in a superconducting 
phase that is very sensitive to the chemical potential, leading to a narrow dome. In the inset to Fig.~3, 
we show results for a larger $g=280$ meV value. In it, we see a much broader variation in electron filling, 
from $0.2$ to $0.31$ [(black) open diamonds curve]. The actual system is quasi-1d, 
implying that once a three-dimensional superconducting state 
stabilizes, it will be less sensitive to variations in the chemical potential. To have the same effect 
in a purely 1d model we have to increase the pairing coupling, as shown in the inset to Fig.~3. 

It is reasonable to expect that applying hydrostatic pressure 
in a crystal lattice will enhance the overlap between the orbitals and therefore increase the magnitude of 
the hopping terms. Taking the reasonable assumption that this increase is similar to the change 
in lattice parameter, which for an applied pressure of 2~GPa will amount to a change 
of $\approx 1\%$ \cite{Tomita2014}, we solve the gap equations for increasing 
values (in magnitude) of $t_{sp}$ and plot the results in Fig.~4 for some $\mu$ values in the dome region 
in Fig.~3 (for $g=81$ meV). For a variation of $|t_{sp}| \approx 0.5\%$ the value of $\Delta$ roughly doubles, 
which is in semi-quantitative agreement with experimental results for $T_c$ obtained for $\rm LaO_{0.5}F_{0.5}BiS_2$ 
and $\rm CeO_{0.5}F_{0.5}BiS_2$ \cite{Wolowiec2013}. A similar calculation for the variation in $t_{pp}$ (not shown) 
shows no changes in $\Delta$, up to the same percent variation as for $t_{sp}$. This seems to be consistent 
with previous results \cite{Continentino2014} showing that antisymmetric hybridization is very effective 
in increasing the SC gap amplitude. To test this hypothesis, extensive calculations are underway where the condition  
$ g_{ss} = g_{p_{b}p_{b}} = g_{sp_{b}} = g $ is relaxed \cite{Griffi2015}. 

There is another choice of pair-scattering terms in eq.~(4) which 
produces results (not shown) very similar to the ones just described. One just needs to replace $s$ by $p_a$ 
in eq.~(4). As already mentioned, these two were the only situations where the results obtained 
were compatible with the criteria described above for acceptance of the gap equation results. 
For all the other possibilities, either the coupling parameter $g$ was unrealistically large or 
the electron-filling was too far removed from 1/8-filling.

\begin{figure}
\centering
\includegraphics[width=3.5in]{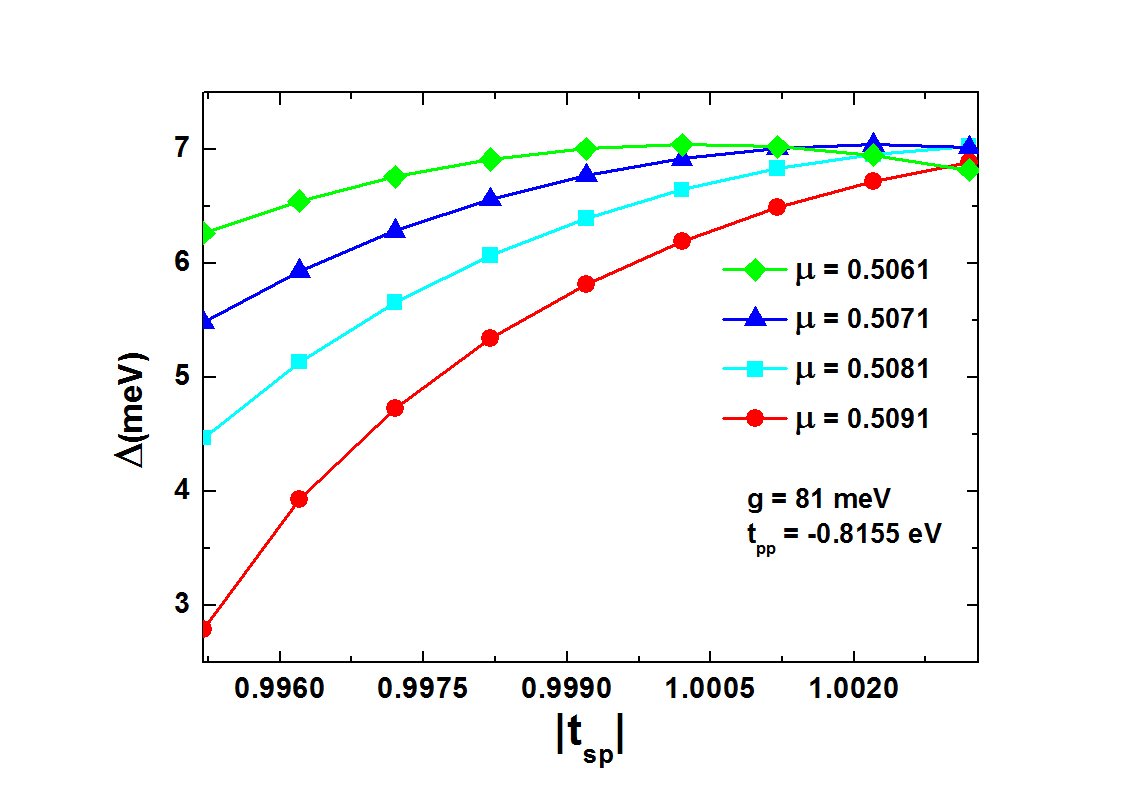}
\caption{Results showing the dependence of $\Delta$ with $t_{sp}$ for a few values of 
	the chemical potential $\mu$ in the dome region in Fig.~3 for $g=81$ meV and $t_{p_ap_b}=-0.8155eV$. 
	The overall variation in the magnitude of $t_{sp}$ is $\approx 0.4\%$, which is a typical 
	lattice parameter variation under typical hydrostatic pressure experiments. 
}
\end{figure}

{\it Conclusions.}
Motivated by recent experiments in superconducting members of the ${\rm BiS_2}$ family of compounds 
showing its `hidden' 1d electronic structure and the strong effect that pressure has 
over its superconducting state, we propose an effective 1d model where the kinetic energy 
part of the Hamiltonian is obtained through DFT calculations for the 2d model for 
${\rm BiS_2}$. Supported by the DOS results shown in Fig.~2, we add the Sulfur p- and s-orbital 
to the p-orbital of Bismuth. Despite being several eV below the other two orbitals, the s-orbital 
undergoes strong hybridization with the Bismuth p-orbital and has a sizable contribution to the DOS at the 
Fermi energy, justifying its inclusion in the model (see Figs.~1 and 2). Pair scattering terms are then added and 
treated at the mean-field level. We solve the gap equations and 
systematically probe what combination of pair-scattering terms produce results in qualitative agreement 
with the experiments, i.e., approximate location of the superconducting phase in a $T$ vs. doping 
phase diagram, realistic coupling constant values, and dependence with hopping parameters (simulating 
application of hydrostatic pressure). We find that single-gap SC 
does not produce acceptable results. This is quite relevant, as 
there is experimental evidence that ${\rm BiS_2}$ presents two gaps \cite{Yazici2015}. 
We find that if we consider $s$- and $p_b$-type pairs, and allow for 
intra and interband scattering we obtain results in semi-quantitative agreement with experiments. The same is true 
if we choose $p_a$- and $p_b$-type pairs, and also allow for intra and interband scattering. The 
interesting point here is that the $t_{sp}$ hopping is the one that, in both cases, enhances SC  
when its magnitude increases, whereas the effect on $\Delta$ of increasing $t_{pp}$ is marginal. This last 
point reinforces the need for considering the Sulfur $s$ orbital explicitly. We argue that the 
anti-symmetric character of the $t_{sp}$ hopping (as stressed in previous work by one of the authors \cite{Continentino2014}) 
may explain its enhanced effect in the superconducting state. 

{\it Acknowledgment.} MAC acknowledges {\it Conselho 
Nacional de Desenvolvimento Cient\'{\i}fico e Tecnol\'ogico - CNPq} and
{\it Funda\c{c}\~ao de Amparo a Pesquisa do Estado do Rio de Janeiro - FAPERJ} 
for partial financial support; KF acknowledges...; MAG acknowledges financial 
support from {\it Conselho Nacional de Desenvolvimento Cient\'{\i}fico e Tecnol\'ogico - CNPq}; and GBM acknowledges 
the Brazilian Government for financial support through a {\it Pesquisador Visitante Especial} 
grant from the {\it Ci\^encias Sem Fronteiras} Program, from the {\it Minist\'erio da Ci\^encia, 
Tecnologia e Inova\c{c}\~ao}. 

\bibliography{thispaper,Bis2-paper,Bis2-experiment,Bis2-theory,Bis2-review,hightc-reviews}

\setcounter{equation}{0}

\section{Supplemental Material}

The gap equations for $\Delta_{1}$ and $\Delta_{2}$ are given by
\begin{eqnarray}
\Delta_{1}&=& \Delta_{ss}+ \Delta^{'}_{p_{b}p_{b}} \nonumber \\
&=& g_{ss}\sum_{k}\langle c_{\bar{k}-}c_{k+} \rangle +  g_{sp_{b}}\sum_{k}\langle p_{b,\bar{k}-}p_{b,k+} \rangle\nonumber
\end{eqnarray}
and
\begin{eqnarray}
\Delta_{2}&=& \Delta^{'}_{ss}+ \Delta_{p_{b}p_{b}} \nonumber \\
&=& g_{sp_{b}}\sum_{k}\langle c_{\bar{k}-}c_{k+} \rangle +  g_{p_{b}p_{b}}\sum_{k}\langle p_{b,\bar{k}-}p_{b,k+} \rangle, \nonumber
\end{eqnarray}
where $\bar{k}=-k$, and the correlation functions are related to the 
Green's functions (propagators, from now on) $\langle\langle c_{k+};c_{\bar{k}-}\rangle\rangle$ 
and  $\langle\langle p_{bk+};p_{b\bar{k}-}\rangle\rangle$ through the equation 
\begin{eqnarray}
\langle \gamma_{\bar{k}-}\gamma_{k+} \rangle &=& i\int^{+ \infty}_{- \infty} d\omega f(\omega)
\lbrack\langle\langle \gamma_{k+};\gamma_{\bar{k}-}\rangle\rangle_{\omega+i\eta}  \nonumber \\
&-&\langle\langle \gamma_{k+};\gamma_{\bar{k}-}\rangle\rangle_{\omega-i\eta}\rbrack, 
\end{eqnarray}
where $\gamma$ stands for the annihilation operators $c$ or $p_{b}$, and $\eta\rightarrow 0$.
In order to calculate the propagators we will write their 
equations of motion (taking from now on $\hbar=1$)
\begin{eqnarray}
\omega \langle\langle c_{k+};c_{\bar{k}-}\rangle\rangle  &=& \frac{1}{2\pi}\langle\{c_{k+},c_{\bar{k}-}\}\rangle \\ \nonumber
&+&\langle\langle[c_{k+},H];c_{\bar{k}-}\rangle\rangle
\end{eqnarray}
and
\begin{eqnarray}
\omega \langle\langle p_{bk+};p_{b\bar{k}-}\rangle\rangle  &=& \frac{1}{2\pi}\langle\{p_{bk+},p_{b\bar{k}-}\}\rangle \\ \nonumber
&+&\langle\langle[p_{bk+},H];p_{b \bar{k}-}\rangle\rangle,
\end{eqnarray}
where $\mathcal{H}(k)$ is the Hamiltonian for the system (eq.~(2) in the main text) 
and $\{,\}$ and $[,]$ indicate an anticommutator and a commutator, respectively. 

Let us develop further the equation of motion 
for the first propagator ($\langle\langle c_{k+};c_{\bar{k}-}\rangle\rangle$). 
Making use of standard relations for fermion creation and 
annihilation operators, we obtain 
\begin{eqnarray}
\omega^{(-)}_{s}\langle\langle c_{k+};c_{\bar{k}-}\rangle\rangle&-&2it_{sp}\sin(k)\langle\langle p_{bk+};c_{\bar{k}-}\rangle\rangle \\ \nonumber
&+&\Delta_{1}\langle\langle c^{\dag}_{\bar{k}-} ;c_{\bar{k}-}\rangle\rangle =0.
\end{eqnarray}

In the process above, two new propagators were created, 
$\langle\langle p_{bk+};c_{\bar{k}-}\rangle\rangle$ 
and $\langle\langle c^{\dag}_{\bar{k}-} ;c_{\bar{k}-}\rangle\rangle$.
In order to close the system of equations for the propagators, we need also the 
equation of motion for $\langle\langle c^{\dag}_{\bar{k}-} ;c_{\bar{k}-}\rangle\rangle$, 
$\langle\langle p_{bk+};c_{\bar{k}-}\rangle\rangle$, 
$\langle\langle p_{b\bar{k}-}^{\dag};c_{\bar{k}-}\rangle\rangle$,  
$\langle\langle p_{ak+};c_{\bar{k}-}\rangle\rangle$, and 
$\langle\langle p_{a\bar{k}-}^{\dag};c_{\bar{k}-}\rangle\rangle$. 
This procedure generates a system of equations given by
\begin{equation}
 D \cdot \left (
 \begin{array}
 [c]{c}%
 \langle\langle c_{\bar{k}-}^{\dag};c_{\bar{k}-} \rangle\rangle \\
 \langle\langle p_{ak+};c_{\bar{k}-} \rangle\rangle \\
 \langle\langle p_{a\bar{k}-}^{\dag};c_{\bar{k}-} \rangle\rangle \\
 \langle\langle c_{k+};c_{\bar{k}-} \rangle\rangle \\
 \langle\langle p_{bk+};c_{\bar{k}-} \rangle\rangle \\
 \langle\langle p_{b\bar{k}-}^{\dag};c_{\bar{k}-}\rangle\rangle
\end{array}
\right )= \left (
 \begin{array}
 [c]{c}%
 \frac{1}{2\pi} \\
 0 \\
 0 \\
 0 \\
 0 \\
 0
\end{array}
\right ), 
\end{equation}
where 
\begin{eqnarray}
D=  \left[
\begin{array}{cccccc}
\omega^{(-)}_{s} & 0& 0 & \Delta_{1}& -\bar{t}_{sp_{b}} &0  \\
0& \omega^{(+)}_{p_{a}}& 0 &  0& 0 & \bar{t}_{p_{a}p_{b}} \\
0 & 0 & \omega^{(-)}_{p_{b}} & 0& -\bar{t}_{p_{a}p_{b}} &0\\
\Delta_{1}& 0& 0 &  \omega^{(+)}_{s}&0 &\bar{t}_{sp_{b}} \\
\bar{t}_{sp_{b}} & 0 & -\bar{t}_{p_{a}p_{b}}  & 0& \omega^{(+)}_{p_{b}} &\Delta_{2}\\
0& \bar{t}_{p_{a}p_{b}}& 0 &  -\bar{t}_{sp_{b}}&\Delta_{2} &\omega^{(-)}_{p_{b}}
\end{array} \right],
\end{eqnarray}
and $\omega^{\pm}_{q}=\omega\pm \epsilon_{q} \mp \mu$, $q=s,p_{a},p_{b}$. 
Here, $\bar{t}_{sp_{b}}=2 i t_{sp_{b}} \sin(k)$ and $\bar{t}_{p_{a}p_{b}}=2 t_{p_{a}p_{b}}\cos(k)$.

Using Cramer's method to solve the system of equations in (5), we have that
\begin{eqnarray}
\langle\langle c_{k+};c_{\bar{k}-}\rangle\rangle = \frac{|D_{ss}|}{|D|}
\end{eqnarray}
where the matrix $D_{ss}$ is obtained by exchanging the 4$^{\rm th}$ column in matrix
$D$ by the column matrix defined in the  right side of eq.~(5) 
(note that $|D|$ means the determinant of matrix $D$).
Repeating the same procedure for $\langle\langle p_{bk+};p_{b\bar{k}-}\rangle\rangle$ we obtain 
\begin{equation}
 D \cdot \left (
 \begin{array}
 [c]{c}%
 \langle\langle c_{\bar{k}-}^{\dag};p_{b\bar{k}-} \rangle\rangle \\
 \langle\langle p_{ak+};p_{b\bar{k}-} \rangle\rangle \\
 \langle\langle p_{a\bar{k}-}^{\dag};p_{b\bar{k}-} \rangle\rangle \\
 \langle\langle c_{k+};p_{b\bar{k}-} \rangle\rangle \\
 \langle\langle p_{bk+};p_{b\bar{k}-} \rangle\rangle \\
 \langle\langle p_{b\bar{k}-}^{\dag};p_{b\bar{k}-}\rangle\rangle
\end{array}
\right )= \left (
 \begin{array}
 [c]{c}%
 0 \\
 0 \\
 0 \\
 0 \\
 0 \\
 \frac{1}{2\pi}
\end{array}
\right )
\end{equation}
where
\begin{eqnarray}
\langle\langle p_{-k};p_{k}\rangle\rangle = \frac{|D_{p_{b}p_{b}}|}{|D|}.
\end{eqnarray}
and $D_{p_{b}p_{b}}$ is obtained by exchanging the 5$^{\rm th}$ column in matrix
$D$ by the column matrix defined in the right side of eq.~(8). 
In eqs.~(7) and (9), $|D|$ is a biquadratic polynomial of degree six and can be rewriten as
\begin{eqnarray}
|D|&=&\sum_{n=0}^{3}B_{2n} \omega^{2n}=\prod_{n=1}^{6}(\omega-\omega_{n}) \\ \nonumber
&=&\prod_{n=1}^{3}(\omega^{2}-\omega^{2}_{n})
\end{eqnarray}
where the last equation is obtained by noting that $\omega_{1}=-\omega_{4}$, 
$\omega_{2}=-\omega_{5}$, and $\omega_{3}=-\omega_{6}$. Here, $A_{n}$ and $B_{n}$ are 
coefficients which are functions of the parameters of the Hamiltonian. 
$\omega_{n}$ are the zeros of $|D|$ and represent the energy excitations of the system. 
The solutions for $\omega_{k}\equiv\omega$ cannot be found analytically. 
Actually, these quantities will be obtained numerically.  

To finally determine the gap equations, it is appropriate to make use of the following identity
\begin{eqnarray}
\frac{1}{|D|}
&=&\frac{1}{r_{1}}\left[\frac{1}{2\omega_{1}} \left( \frac{1}{\omega-\omega_{1}}-\frac{1}{\omega+\omega_{1}}\right) \right ] \\ \nonumber
&+&\frac{1}{r_{2}}\left[\frac{1}{2\omega_{2}} \left( \frac{1}{\omega-\omega_{2}}-\frac{1}{\omega+\omega_{2}}\right ) \right ] \\ \nonumber
&+&\frac{1}{r_{3}}\left [\frac{1}{2\omega_{3}} \left ( \frac{1}{\omega-\omega_{3}}-\frac{1}{\omega+\omega_{3}}\right )  \right],
\end{eqnarray}
where
\begin{eqnarray} \nonumber
r_{1}&=&(\omega^{2}_{1}-\omega^{2}_{2})(\omega^{2}_{1}-\omega^{2}_{3})\\ \nonumber
r_{2}&=&(\omega^{2}_{2}-\omega^{2}_{1})(\omega^{2}_{2}-\omega^{2}_{3}) \\ \nonumber
r_{3}&=&(\omega^{2}_{3}-\omega^{2}_{1})(\omega^{2}_{3}-\omega^{2}_{2}). \nonumber
\end{eqnarray}

Substituting eq.~(11) into eqs.~(7) and (9), and after using eq.~(1), we have
\begin{eqnarray}
\langle \gamma_{-k}\gamma_{k} \rangle
&=&i\sum^{3}_{j=1}\int^{+ \infty}_{- \infty}d\omega 
f(\omega)\frac{D_{\gamma \gamma}(\omega)}{2\omega_{j}r_{j}} \\ \nonumber 
&\times& (C_+(\omega)-C_-(\omega)),
\end{eqnarray}
where we have defined the following quantities
\begin{eqnarray}
C_{\pm}(\omega) = \lim_{ \eta \rightarrow 0^{+}}(\frac{1}{\omega \pm \omega_{j}-i\eta}-\frac{1}{\omega \pm \omega_{j}+i\eta}).
\end{eqnarray}
Now, using the fact that
\begin{eqnarray}
\delta(x)=\frac{1}{2\pi i}\lim_{ \eta \rightarrow 0^{+}}\left(\frac{1}{x-i\eta}-\frac{1}{x+i\eta}\right),
\end{eqnarray}
we finally get
\begin{eqnarray}
\langle \gamma_{-k}\gamma_{k} \rangle &=& \pi \sum^{3}_{j=1} \frac{ |D_{\gamma \gamma}(\omega_{j})| f_{FD}(\omega_{j}) }{ \omega_{j}r_{j} } \\ \nonumber
&-&\pi \sum^{3}_{j=1} \frac{ |D_{\gamma \gamma}(-\omega_{j})| f_{FD}(-\omega_{j}) }{ \omega_{j}r_{j} },
\end{eqnarray}
where $f_{FD}(\omega_{j})=\frac{1}{\exp(\omega_{j}/k_{B}T)+1}$.

Substituting eq.~(15) into the equations for $\Delta_{1}$ and $\Delta_{2}$ in the 
previous page, and taking the limit $T\rightarrow0$, we can write the following self-consistent equations
\begin{eqnarray}
\Delta_{1}&=& \Delta_{ss}+ \Delta^{'}_{p_{b}p_{b}} \nonumber \\
&=& -g_{ss}\sum_{k}\pi \sum^{3}_{j=1} \frac{ |D_{ss}(-\omega_{j})| }{ \omega_{j}r_{j} } \nonumber \\
&-&  g_{sp_{b}}\sum_{k}\pi \sum^{3}_{j=1} \frac{ |D_{p_{b}p_{b}}(-\omega_{j})|}{ \omega_{j}r_{j} }
\end{eqnarray}
and
\begin{eqnarray}
\Delta_{2}&=& \Delta^{'}_{ss}+ \Delta_{p_{b}p_{b}} \nonumber \\
&=& - g_{sp_{b}}\sum_{k}\pi \sum^{3}_{j=1} \frac{ |D_{ss}(-\omega_{j})| }{ \omega_{j}r_{j} } \nonumber \\
&-&  g_{p_{b}p_{b}}\sum_{k}\pi \sum^{3}_{j=1} \frac{ |D_{p_{b}p_{b}}(-\omega_{j})|}{ \omega_{j}r_{j} }.
\end{eqnarray}

In the thermodynamic limit ($L\rightarrow \infty $), we can replace the 
sum by an integral [in the interval $-k_{f}\leq k\leq k_{f}$] by using the standard relation 
\begin{eqnarray}
\frac{2\pi}{L}\sum_{k} \Delta_{k}=\frac{1}{2\pi}\int^{ k_{f}}_{-k_{f}} dk,
\end{eqnarray}
where, $k_{f}$ is the Fermi-wavevector and $L$ is 
the length of the one-dimensional system. For practical purposes, we change 
the integration variable ($\tilde{k}=k/k_{f}$), and finally write
\begin{eqnarray}
\Delta_{1}&=& -g_{ss}k_{f} \pi \int^{1}_{-1} d\tilde{k} \sum^{3}_{j=1} \frac{ |D_{ss}(-\omega_{\tilde{k},j})| }{ \omega_{\tilde{k},j}r_{j} } \nonumber \\
&-&  g_{sp_{b}}k_{f} \pi \int^{1}_{-1} d \tilde{k} \sum^{3}_{j=1} \frac{ |D_{p_{b}p_{b}}(-\omega_{\tilde{k},j})| }{ \omega_{\tilde{k},j}r_{j} },
\end{eqnarray}
and
\begin{eqnarray}
\Delta_{2}&=& -g_{sp_{b}}k_{f} \pi \int^{1}_{-1} d \tilde{k} \sum^{3}_{j=1} \frac{ |D_{ss}(-\omega_{\tilde{k},j})| }{ \omega_{\tilde{k},j}r_{j} } \nonumber \\
&-&  g_{p_{b}p_{b}}k_{f} \pi \int^{1}_{-1} d \tilde{k} \sum^{3}_{j=1} \frac{ |D_{p_{b}p_{b}}(-\omega_{\tilde{k},j})| }{ \omega_{\tilde{k},j}r_{j} }.
\end{eqnarray}
In a one-dimensional system, $k_{f}=\frac{\rho\pi}{2}$, 
where $\rho=\frac{N}{L}$ is the density of electrons in the material and $N$ is 
the total number of electrons. Note that, when $g_{ss}=g_{p_{b}p_{b}}=g_{sp_{b}}$, we 
have $\Delta_{1}=\Delta_{2}$.

\end{document}